\documentclass[twocolumn]{aastex61}
\received{July 25, 2017}
\revised{September 15, 2017}
\accepted{September 19, 2017}
\submitjournal{ApJ}

\shorttitle{Type I shell galaxies as a test of gravity}
\shortauthors{Vakili et al.}

\begin{document}

\title{Type I shell galaxies as a test of gravity models}

\correspondingauthor{Hajar Vakili}
\email{vakili@physics.sharif.edu}

\author{Hajar Vakili}
\affil{Department of Physics, Sharif University of Technology, P.O. Box 11365-9161, Tehran, Iran}

\author{Pavel Kroupa}
\affiliation{Helmholtz-Institut f\"ur Strahlen-und Kernphysik, Universit\"at Bonn, Nussallee 14-16, D-53115 Bonn, Germany}
\affiliation{Charles University in Prague, Faculty of Mathematics and Physics, Astronomical Institute, V Hole\v{s}ovi\v{c}k\'ach 2,~CZ-180 00 Praha 8, Czech Republic}
\nocollaboration

\author{Sohrab Rahvar}
\affiliation{Department of Physics, Sharif University of Technology, P.O.Box 11365-9161, Tehran, Iran}
\nocollaboration



\begin{abstract}

Shell galaxies are understood to form through the collision of a dwarf galaxy with an elliptical galaxy. Shell structures and kinematics have been noted to be independent tools to measure the gravitational potential of the shell galaxies.
We compare theoretically the formation of shells in Type I shell galaxies in different gravity theories in this work because this is so far missing in the literature. We include Newtonian plus dark halo gravity, and two non-Newtonian gravity models, MOG and MOND, in identical initial systems.
We investigate the effect of dynamical friction, which by slowing down the dwarf galaxy in the dark halo models limits the range of shell radii to low values. 
Under the same initial conditions, shells appear on a shorter timescale and over a smaller range of distances in the presence of dark matter than in the corresponding non-Newtonian gravity models. 
If galaxies are embedded in a dark matter halo, then the merging time may be too rapid to allow multi-generation shell formation as required by observed systems because of the large dynamical friction effect.
Starting from the same initial state, in the dark halo model the observation of small bright shells should be accompanied by large faint ones, while for the case of MOG, the next shell generation patterns iterate with a specific time delay.
The first shell generation pattern shows a degeneracy with the age of the shells and in different theories, but the relative distance of the shells and the shell expansion velocity can break this degeneracy.

\end{abstract}




\section{Introduction} \label{sec:intro}

Studying dynamics of giant elliptical galaxies is difficult because we lack kinematic tracers at large radii, where there is almost no way for a direct measurement of the gravitational potential (\citealt{richtler2008dark, samurovic2014investigation, bilek2015mond}).
Shell galaxies are elliptical or S0 galaxies surrounded by faint arc-like structures made of stars.
\cite{merrifield1998measuring} used shell kinematics to measure the potential of shell galaxies.
\cite{ebrova2013shell} used the shell kinematics to reconstruct the parameters of the potential of the shell galaxies from the simulated data and extended its usage as an independent tool to determine the distribution of dark matter in these galaxies up to radii $\approx 100~\mathrm{kpc}$.~\cite{bilek2013testing, bilek2014mond, bilek2015mond} pioneered the field by using these systems as a powerful tool to constrain MOdified Newtonian Dynamics (MOND) on galactic scales.
In addition, observational evidence for the action of dynamical friction that is due to the expansive dark matter halos surrounding galaxies has been suggested as a powerful test for the existence of dark matter~(\citealt{kroupa2014galaxies}).

Shell galaxies were reported for the first time in the Atlas of Peculiar Galaxies by~\cite{arp1966atlas}.~\cite{malin1977unsharp} and~\cite{malin1980giant} discovered shells in a substantial fraction of early-type galaxies by applying new photographic techniques.~\cite{malin1983catalog} published a catalog of 137 shell galaxies in the ESO/SRC Southern Sky Survey. \cite{wilkinson1987two} and~\cite{prieur1990status} categorized shell galaxies into three different morphological types: 
axially symmetric shell systems interleaving in radius are classified as Type I. They are the simplest systems to study analytically.
Type II includes randomly distributed arcs all around a rather circular galaxy. Irregular and complex structure of shells are of Type III. 

The formation process of shells was widely investigated in the 1980s, e.g. by~\cite{schweizer1980optical}, \cite{fabian1980star}, \cite{williams1985blast}, and \cite{thomson1990weak}. Among other proposed formation models, shell systems are dominantly accepted to be remnants of minor mergers of dwarf galaxies (the secondary) with an elliptical (the primary) as the host galaxy. Tidal forces within the host cannibalize the dwarf after the encounter, while the host remains intact. Detached stars from the dwarf generate density waves in the primary and form the shell system. 
Simulations based on the merger model were first made by~\cite{quinn1984formation} and were followed by~\cite{ dupraz1986shells, dupraz1987dynamical}, \cite{hernquist1987shell, hernquist1988formation, hernquist1989formation}, and most recently by, e.g.~\cite{ebrova2012quadruple} and \cite{bilek2014shell}.

In this study, we use the simplified model of the Type I shell system as the consequence of a radial impact of a dwarf galaxy in the fixed spherically symmetric gravitational potential of the massive host galaxy.
In this picture, which is consistent with the abovementioned simulations, the incoming galaxy is released from rest to move radially toward the host galaxy from a certain initial distance.
Its motion can be described as a damped oscillator around the center of the primary, which releases a part of its mass when it passes through the center. The released stars are assumed to oscillate freely in the gravity of the host galaxy,
so the surface brightness increases near the turning points, where the stars spend more time and the shell system is generated.
If the dwarf survives after its first passage, the core remnant continues oscillating, and further passages lead to the next shell generations.
Movement of the core within the massive host is affected by dynamical friction, the overall gravitational effect of the host mass, which gradually slows the motion of the secondary down. Thus, the matter content of the primary plays an important role in the motion of the secondary and for the array of the next generations of shells. In modified gravity theories, the effect of dynamical friction is less important because in comparison with the dark matter models, the amount of matter is much smaller.

The aim of this study is to investigate the characteristics of shell patterns and possible differences between them in different gravity theories for the identical initial systems because a study like this is so far missing in the literatue. We verify the power of shell kinematics in model distinction and testing the fundamental physics. 
In this work, shell formation in the standard dark matter scenario is compared with the two alternatives to the dark 
matter paradigm: (a) MOdified Gravity (hereafter MOG) as a modification to the General Relativity (GR), and (b) MOND.
The paper organization is as follows. In Section \ref{sec:models} we briefly introduce the gravity theories that we compare. The theory of shell formation and expansion is discussed in Section \ref{sec:shell_theory}. 
In Section \ref{sec:shells_gravities} we compare the motion of the incoming galaxy and the resulting shells in dark matter or modified gravity theories  and introduce some probes that can be used to distinguish between various gravitational models.
Our conclusion appears in Section \ref{sec:conclusion}.

\section{Models of the dynamics of galaxies}\label{sec:models}

Dark matter haloes within and around the galaxies are introduced to interpret the rotation curves of spiral galaxies and velocity dispersion of elliptical galaxies~\citep{bertone2005particle}.
On the scales of galaxies, there are alternative models of modified gravity, such as MOG and MOND, and also nonlocal gravity models 
\citep{hehl2009nonlocal, Blome:2010xn, mashhoon2017nonlocal} to explain the dynamics of galaxies without dark matter. 
Rotation curves of many disk galaxies~\citep{haghi2016declining} including solar system constraints~\citep{hees2015combined} have been studied to test the MOND theory.
In this paper, we study the formation of shell galaxies, which are suitable systems for observational tests of modified gravity in comparison with dark matter models.  
In this section, we introduce the model of the dark matter halo of galaxies and its alternatives, MOG and MOND.

\subsection{Dark Matter Halo}\label{subsec:dm_halo}

The host galaxy, a typical elliptical, is modeled as a luminous homogenous sphere of $5~\mathrm{kpc}$ radius, and the dark matter contribution to the potential is modeled with a spherical virialized halo surrounding the stellar mass of the primary.
We choose the $\mathrm{NFW}$ density profile that \citet{navarro1997universal} proposed for describing the cold dark matter (CDM) haloes of galactic structures (see also \citealt{kroupa2010local}). Given the critical density of the Universe, $\rho _0$, the halo density at distance $r$ from the center is defined by
\begin{equation}
\label{eq:NFW}
\rho_{\rm{NFW}}(r) =\frac{\delta _{\rm c}\rho _0}{r/r_{\rm{s}}\left(
  1+r/r_{\rm{s}}\right) ^2},
\end{equation}
where the characteristic radius $r_{\rm{s}}$ and the parameter $\delta_{\rm c}$ are defined by the total mass of the halo.
The mass within a certain radius is obtained by integrating the density over the volume,
\begin{equation}
M(r) ={4 \pi \rho_0 \delta _c r_{\rm s}^3} \; \left[
  \frac{1}{1+\;r/r_{\rm{s}}}+\ln \left(
  1+\;r/r_{\rm{s}} \right) -1\right].
\label{eq:enclmass2}
\end{equation}
The virial mass of the halo, the total mass within the virial radius, $r_{\rm vir}$, is defined as
\begin{equation} 
M_{\rm{vir}}=\frac{4 \pi}{3} \Delta_{\rm{vir}}\rho_{0} r_{\rm{vir}}^3,
\label{eq:VirMass}
\end{equation}
where $\Delta_{\rm{vir}} \rho_{0}$ is equal to the critical density at which matter overcomes the cosmic expansion and collapses into a virialized halo.
The value of $\Delta_{\rm{vir}}$ for the local universe (redshift equal to zero) is approximately equal to $104.2$ for $\mathrm {\Lambda DCM}$ cosmology according to \citet{dutton2014cold}. Introducing the concentration parameter $c_{\rm{vir}}=r_{\rm{vir}}/r_{\rm{s}}$ and setting $r=r_{\rm{vir}}$ in Equation (\ref{eq:enclmass2}), the parameter $\delta _{\rm c}$ is obtained as
\begin{equation}
\delta _{\rm c} =\frac{\Delta_{\rm{vir}}}{3} \frac{c_{\rm{vir}}^3}{\ln \left( 1+c_{\rm{vir}}  \right) -c_{\rm{vir}}/(1+c_{\rm{vir}})}.
\label {eq:delta}
\end{equation}
The relation between $c_{\rm{vir}}$ and $M_{\rm{vir}}$ in the local universe is obtained from simulations \citep{dutton2014cold} as
\begin{equation}
\log_{10}(c_{\rm{vir}}) = 1.025-0.097 \log_{10} \left(\frac{M_{\rm{vir}}} {10^{12}h^{-1}{ M_{\odot}}}\right).
\label {eq:C_M_relation}
\end{equation}

To achieve the goal of this study, which is to compare the theories under the same conditions, we completely focus on the theoretical expectations of each model, instead of using observational data.
To model the dark matter halo of an elliptical galaxy, we use the Illustris cosmological simulation, which is based on $\rm{\Lambda CDM}$ Cosmology.
The contribution of the total baryonic mass in this simulation is derived from Figure 1 of ~\citet{haider2016large}.
Using the curve labeled as "Illustris, total baryons," for a structure with a given total halo mass, one can define the baryonic fraction of the structure. We choose two different masses for the dark matter halo to model the host galaxy.
The characteristics of the models are described in Table \ref{tab:models}.

\begin{deluxetable}{lcccc} 
\tablecaption{Dark and luminous matter properties of the model elliptical galaxy. \label{tab:models}}
\tablehead{
\colhead{ Models} & \colhead{$M_{\rm{b}}$ } & \colhead{$r_{\rm{b}}$ } &  \colhead{ $M_{\rm{vir}}$ } & \colhead{$r_{\rm{vir}}$ }\\
\colhead{} & \colhead{$\left({\rm M_{\odot}}\right)$} & \colhead{$(\rm{kpc})$} & \colhead{$\left({\rm M_{\odot}}\right)$} & \colhead{$\left({\rm kpc}\right)$}
}
\startdata
		Halo model A  & $5\times10^{11} $ & 5 & $10^{13} $ & 568 \\
		Halo model B & $10^{11}$ 	      & 5 & $10^{12} $ & 264\\
		MOG	      & $10^{11}$	      & 5 &...&...\\
		MOND	      & $10^{11}$	      & 5 &...&...\\
\enddata
\tablecomments{$M_{\rm{b}}$ and $r_{\rm{b}}$ are the total baryonic mass and radius, respectively. $M_{\rm{vir}}$ is the total halo mass within $r_{\rm{vir}}$, the virial radius of the dark halo.}
\end{deluxetable}

\subsection{Modified Gravity (MOG)}

MOG has been proposed by~\cite{moffat2006scalar} to explain galactic dynamics using the existing baryonic matter. Being a covariant extension of GR, this model is derived from the action principle that introduces two scalar fields and a vector field in addition to GR fields. The effective potential in the weak-field approximation of MOG is a combination of a strong Newtonian-like attraction and a Yukawa-like repulsive term with two additional parameters $\alpha$ and $\mu$~\citep{moffat2013mog}.

For a nonrelativistic test-mass particle in the distribution of matter $\rho(\textbf{\textit{r}})$,  the gravitational acceleration 
\textbf{\textit{a}}(\textbf{\textit{r}}) is equal to
\begin{eqnarray}\label{eq:potential22}
\textbf{\textit{a}}(\textbf{\textit{r}})  &=&- {G}\int_0^R \rho(\textbf{\textit{r}}')\frac{(\textbf{\textit{r}} - \textbf{\textit{r}}')} {|\textbf{\textit{r}}-\textbf{\textit{r}}'|^3} \times \\
&& \left\{ 1+\alpha- \alpha \rm e^{-\mu|(\textbf{\textit{r}} - \textbf{\textit{r}}')|} \left( 1+\mu|(\textbf{\textit{r}} - \textbf{\textit{r}}')| \right) \right \} d^3\textbf{\textit{r}}'. \nonumber 
\end{eqnarray}
For large scales compared to $\mu^{-1}$ the gravity is $(1+\alpha)$ times stronger than the Newtonian case, while for 
small scales, Newtonian dynamics is fully established.

This model has been successful in describing galactic rotation curves. \cite{moffat2013mog} fit theoretical rotation curves of galaxies to observational data and found the best-fit values for the free parameters to be $\alpha = 8.89 \pm 0.34$ and $\mu= 0.042\pm 0.004~\mathrm{kpc}^{-1}$. We use the same values in this paper as the universal model parameters. 
In addition, MOG has been shown to be consistent with ionized gas and temperature profiles of nearby clusters obtained by the {\it Chandra} X-ray telescope \citep{moffat2014mog}. 

\subsection{Modified Newtonian Dynamics}

MOND is another alternative to the non-baryonic dark matter proposed by~\citet{milgrom1983modification}. Based on MOND, gravity or inertia does not follow the prediction of Newtonian dynamics for accelerations smaller than a specified threshold of 
${a}_0 \approx 1.2 \times 10^{-10}~m~s^{-2}$. For the case that the acceleration of a test-mass particle in the gravitational field is 
slower than the characteristic acceleration of the theory, ${a}_0$, the acceleration is related to the Newtonian 
acceleration $a_{\rm N}$ by $a=\sqrt{{a}_0 a_{\rm N}}$. 
This regime is called "deep-MOND". In the intermediate range, the interpolating function $\mu (x)$ describes the transition from the Newtonian gravity to the deep-MONDian regime:
\begin{equation}\label{eq:mond}
a_{\rm N}=a \mu(a/{a}_0),
\end{equation}
where the standard form is
\begin{equation}
\mu (x)=\frac{x}{\sqrt{1+x^2}}.
\end{equation}
The model recovers the Newtonian gravity for large accelerations in the strong-field regime. 

Since its proposal in 1983, MOND succeeded in interpreting many observations without invoking the existence of dark matter. 
The most prominent observational pieces of evidence of MOND are galactic rotation curves, e.g.~\cite{begeman1991extended}, the Milgromain dynamics tests for the space-time scale-invariance~\citep{milgrom2009mond}, the baryonic Tully-Fisher relation, and the mass discrepancy-acceleration relation~\citep{mcgaugh2004mass, wu2015galactic}, or alternatively, the radial acceleration anomaly (\citealt{mcgaugh2016radial, Lelli:2017vgz}). For a major review see \cite{famaey2012modified}.

\section{Theory of shell formation}\label{sec:shell_theory}

The theories of the formation of shell galaxies can be categorized into three classes. The first one includes the gas dynamical theories proposed by~\cite{fabian1980star}, which connect the star formation and the formation of shells. However, these theories are ruled out by observations and are commonly not considered. The second class is the weak-interaction model introduced by~\cite{thomson1990weak}. In this model, a weak interaction of a thick-disk population of dynamically cold stars with another galaxy induces density waves and forms the shells. Although it has nice explanations for shells, this model has some deficiencies and obscurities (see \citealt{ebrova2013shell} for more details on these models). 
Finally, the most widely accepted theories for shell formation are the merger models that came from the idea of~\cite{schweizer1980optical}.
According to these models, shells are the consequence of the encounter of two galaxies. 
The stars that formerly belonged to the smaller galaxy become trapped into the potential of the host galaxy after the collision and start to oscillate freely around the center of the host.
Shells appear near the apocenters of the oscillating stars, where the surface brightness increases as a result of the rise in the number density of the stars.
The details depend on the morphology of the colliding galaxies and on the collision conditions. A nearly radial merger along the major axis of the primary generates shells as parts of spheres that are interleaved in radius. These shell systems are classified as Type I~\citep{hernquist1988formation, hernquist1989formation}. See~\cite{bilek2016galaxy} for a detailed review of the other types. 

The time evolution of the Type I shell system can be described analytically, so they can be used to constrain the potential of the host galaxy.
In general, the first passage of the dwarf releases a fraction of its stars that oscillate freely in the potential of the host, forming the first-generation shells. Subsequent passages of the secondary each time release more stars. The second passage, for example, forms a second-generation shell system. 
As the shells are the locus of the same-phased stars at their turning points, 
\cite{hernquist1987shell} used the relation between the position of the $n$th ordered shell from, e.g., the first generation, $r_n$, at any given time after the encounter with the host gravitational potential, $\phi$, approximately as
\begin{equation}\label{eq:shell_radi}
t(r_n)=\left(n+\frac{1}{2}\right)P(r_n),
\end{equation}
where $n$ starts from $1$ and $P(r_n)$, the half-period of the stellar motion, is given by
\begin{equation}\label{eq:period}
P(r_n)=\sqrt{2}  \int_0^{r_n} {\frac{{d}x}{\sqrt{\phi(r_n)-\phi(x)}}}.
\end{equation}
We use the approximate statement by~\cite{dupraz1986shells} that the velocity of the stars in a shell is equal to the phase velocity of the shell, which is the evolution of the position of each turning point during time. 
Thus, the expansion velocity of a shell is approximately equal to 
\begin{equation}\label{eq:exp_vel}
v_n=\left[ {\left(n+\frac{1}{2}\right)\frac{{d} {P(r_n)}}{{d}{r}}} \right]^{-1}.
\end{equation}
The use of this equation is accurate enough to study shell kinematics according to Table 2 of~\cite{ebrova2012quadruple}.
 
Observations of shells in a shell galaxy exhibit a snapshot of what happens in the shell formation process. 
Equations (\ref{eq:shell_radi})-(\ref{eq:exp_vel}) show that the distance between the shells and the shell expansion velocity depend only on the gravitational potential of the host galaxy.
Thus, the size, separation, and pattern of the shell structure and the shell expansion velocity will differ for different gravitational theories.

\begin{figure}
	\includegraphics[width=1.0\columnwidth]{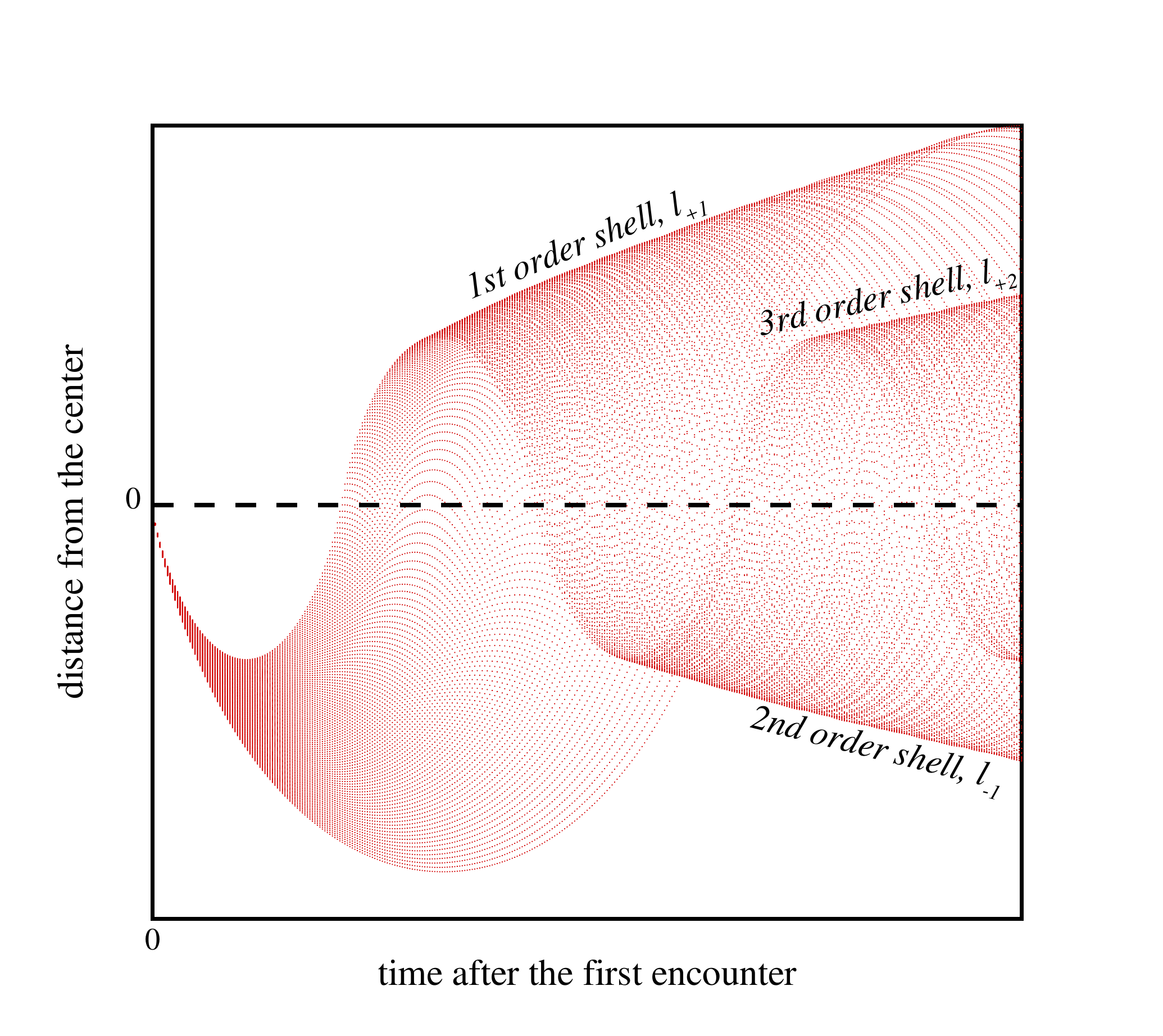}
   \caption{Schematic stellar trajectories depicting shell formation and expansion after the encounter. Dwarf stars leave their host at time 0 and oscillate freely in the elliptical galaxy potential. Near their apocenters, where the stars spend more time, the surface brightness increases and shells of different order are generated. The shells implied here are all from the first-generation system.
        For further discussions see Sections \ref{sec:shell_theory} and \ref{sec:shell_structure}. }
    \label{fig:shell}
\end{figure} 
\begin{figure}
	\includegraphics[width=1.0\columnwidth]{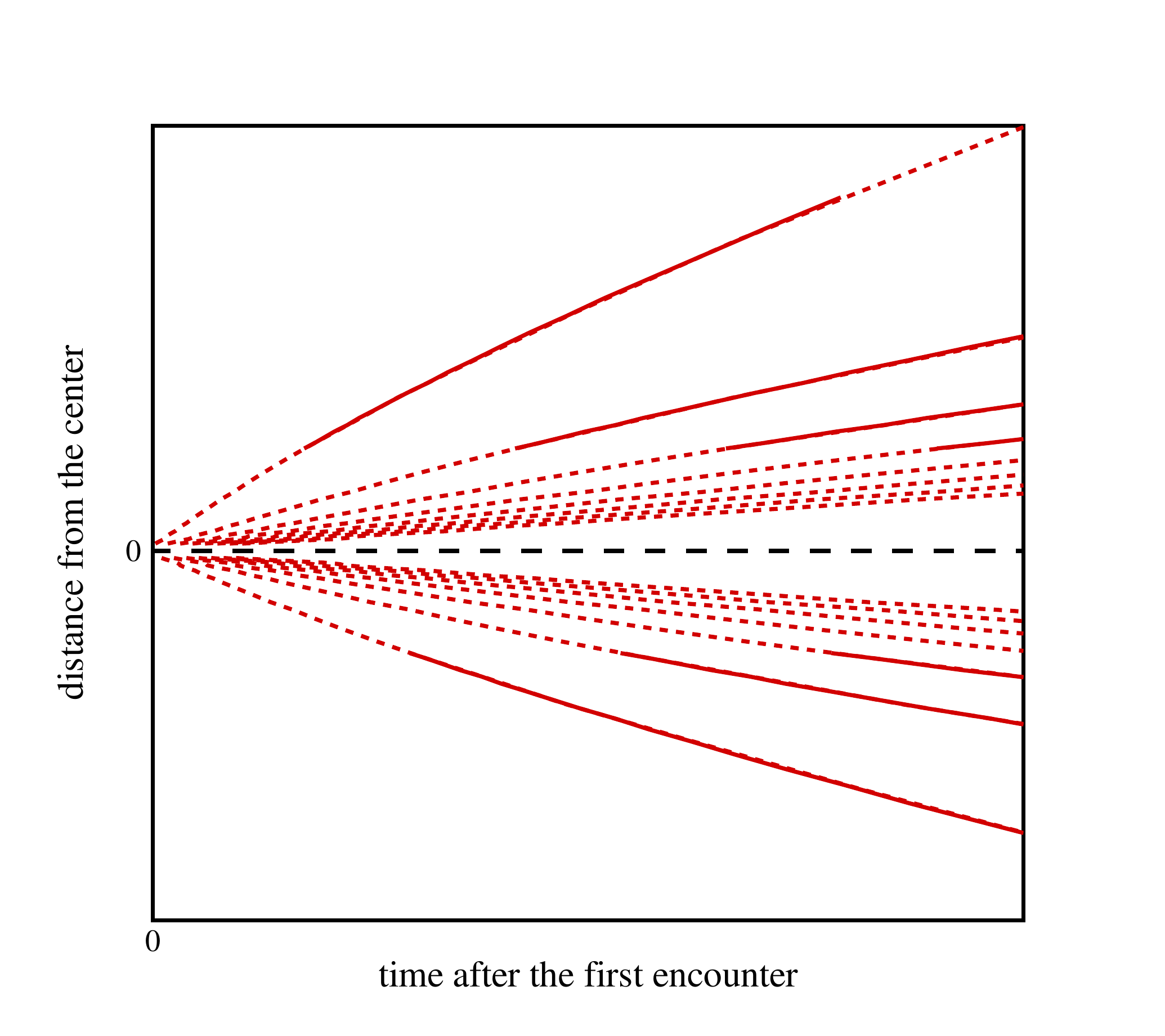}
   \caption{Schematic shell radii after the encounter. Dashed red lines show the theoretically possible shell positions with no limit on stellar kinetic energies. Solid lines are the visible part of the shells considering the distribution of the stars in velocity.
   Thus, the solid line parts correspond to stars with speeds $v \pm \Delta v$, while the dashed segments are shells populated by stars with speeds beyond this interval. 
        For further discussion see Section \ref{sec:shell_br}. }
    \label{fig:shell_vel}
\end{figure} 

Following the stellar trajectories in the primary's gravitational potential after they left the secondary, we determine the shell structure generated by the collision. 
Assuming the Newtonian gravity (halo model A in Table \ref{tab:models}), Figure~\ref{fig:shell} schematically visualizes the dwarf star trajectories in the elliptical galaxy gravity and the resulting shell system for the dark matter halo model.
The process of shell formation and its properties is clearly shown in this figure. It can be seen that 
the concentration of stars increases near the apocenters, where the shells are formed. 
The star-orbit turning points do not line up in the first oscillation and the first shell appears in the next peak, i.e. in the direction from which the dwarf galaxy came.
As the oscillation continues, the slowest stars that already formed the first shell leave it to make the second shell, and the more initially rapid stars replace them. This leads to the apparent shell expansion that is approximately described by Equation (\ref{eq:exp_vel}). Measuring the slope of the shell envelopes in this figure yields the expansion velocity. The outermost shell at any time consists of the most rapid stars that already completed three-fourths of their first orbit, and the innermost shell represents the initially slowest stars with the smallest oscillation amplitude.
Based on this figure, it can be seen that all of the properties of the shells can be understood using the idealized condition that we used.

\subsection{Shell Brightness and Velocity}\label{sec:shell_br}

The faster that any star is when it leaves its host dwarf galaxy (the socondary), the farther it moves away, and the larger the shell it contributes to form.
Thus, the surface brightness of a shell, determined by the number density of the stars in that shell, depends on the initial velocity distribution of its stars.
When we assume the initial velocity of the released stars to have a Maxwellian distribution around the velocity of the secondary at the time of passage through the center of the host ($v$), the stellar speeds mainly lie in the range $v \pm \Delta v$, considering the velocity dispersion they had in the dwarf before leaving it ($\Delta v$). This will limit the visibility of the large shells because of the small number density of the stars beyond this range. 

Our goal is to find the positions where shell formation is possible.
Thus, without loosing generality, one can assume that the energies of the released stars cover a continuous interval so that
the motion of the stars poses all the possible velocities between zero and infinity
and the shells live forever once they are generated after a passage.
This assumption is compatible with the statement by \cite{bilek2013testing} and their Figure 3.
As in the case of the limited velocity distribution for the stars, in this case, one has to keep in mind that the surface brightness of the shells should vary in time because the initial velocity distribution of the stars is not uniform.  
According to simulations, the surface brightness profile of the shells depends on the original size and velocity dispersion of the secondary in addition to the potential of the primary~\citep{hernquist1988formation}. 
Figure~\ref{fig:shell_vel} schematically shows the effect of the limitaing the initial stellar velocities in the visibility of shells. The large shells will fade after a long time and escape from observational accessibility, although it is theoretically possible for them to occur. Moreover, the very small shells would not be visible owing to the small number density of the stars.
Studying the surface brightness profile of the shells in a quantified manner is beyond the scope of this paper. 

\section{Shell galaxies in different gravities}\label{sec:shells_gravities}

The characteristics of the shell system that follows after the encounter of two galaxies depend on the properties of the collision. For an idealized study, four elements are effective in the final shell pattern: the model of the gravitational potential, the mass ratio between the two galaxies, the mass-loss ratio of the secondary in each passage, and the initial velocity or equivalently the initial distance at which the secondary starts its motion from the rest velosity.

The merger time-scale and the relative encounter velocity depend on the initial distance between the two galaxies. Moreover, dynamical friction permits the mass ratio between the elliptical and the dwarf to play an important role in the formation of shells, especially in the presence of a dark matter halo.
It is hard to deal with dynamical friction without any numerical simulation. 
While there are \textit{N}-body codes for MOND and LCDM that enables a more realistic treatment with dynamical friction, e.g., \citet{teyssier2002cosmological}, \citet{lughausen2014phantom}, and \citet{oehm2017constraints}, 
a complete comparison between the models with the same method is not feasible until we have such codes for MOG. 
Meanwhile, it is possible to do the comparison analytically using the Chandrasekhar formulation in the symmetric conditions of our study, 
which is described in the Appendix, and its validity in galactic dynamics is verified by simulations.
In addition, the type of shell galaxy or the shape of the shell pattern are directly related to the impact angle. 
Type I shell galaxies result from nearly radial mergers~\citep{hernquist1988formation}, thus it is reasonable to assume head-on collisions to avoid non-symmetric shells. This approximation is furthermore useful because we aim to distill the differences in shell patterns under different description of gravitation. 

The most decisive factor is the gravitational potential of the host galaxy that governs the motion of the secondary and its stars within the host.
The contribution of gravity cannot be determined explicitly, unless when we have the initial condition of the impact under control. In this section, we therefore follow the effect of the initial conditions on the motion of the secondary until the end of merging, as well as their effects on the shape of the shells.

\subsection{A comparison between the models}\label{sec:acc}

\begin{figure}
\includegraphics[width=1.05\columnwidth]{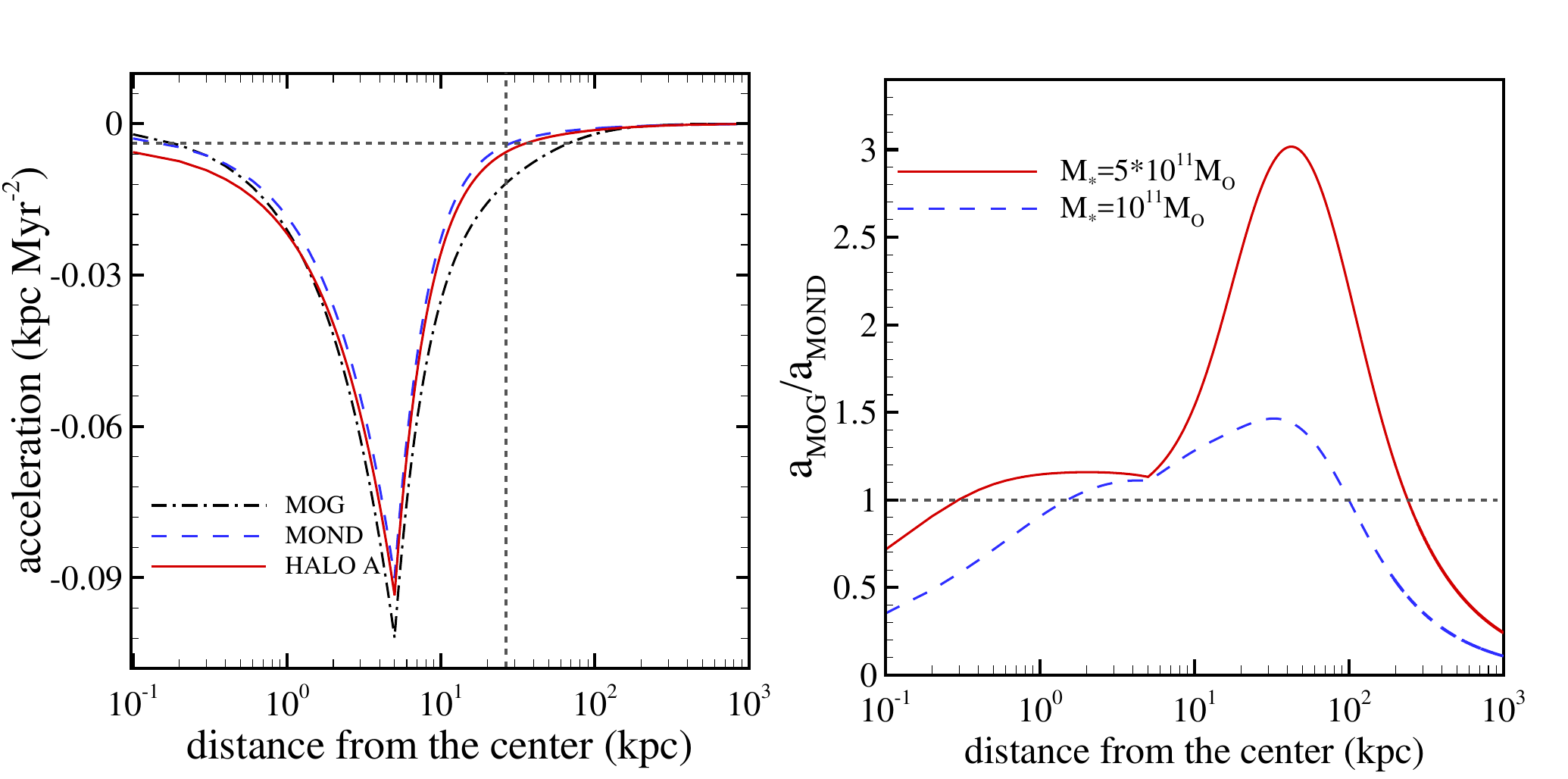}
   \caption{Left: acceleration of a test particle in different models in terms of distance from the center of a $5~\rm kpc$ uniform sphere. Here the mass of the luminous sphere is considered to be the same as in the halo model A, \textit{i.e.} $5\times 10^{11}{\rm M}_{\odot}$. The halo model A is shown with the red solid line (see the Table~\ref{tab:models}), MOG with the black dash-dotted line and MOND with the blue dashed line. The horizontal dashed line indicates the MOND threshold for acceleration, ${a}_0$, and the vertical line is the length scale of MOG $\approx 24~\rm kpc$. 
   Right: ratio of the acceleration of a test particle in MOG over MOND in the gravitational field of two uniformly distributed mass spheres each with a radius of $5~\rm kpc$. The red solid line is calculated for a total baryonic mass of $5 \times 10^{11}{\rm M}_{\odot}$ ($r_0=24.33\rm~kpc$ in Equation (\ref{eq:MONDr0})) and the blue dashed line is related to a $10^{11}{\rm M}_{\odot}$ mass ($r_0=10.88\rm~kpc$ in Equation (\ref{eq:MONDr0})).}
    \label{fig:acc_ratio}
\end{figure}

According to Equation (\ref{eq:period}) the relevant factor that determines the position of the shells is the gravitational field of the host galaxy, or equivalently, the acceleration of the particles in that field.
Prior to the analysis of the shell formation mechanism in different models, it is useful to have a qualitative comparison between the accelerations that the secondary, or its liberated stars, undergo on their trajectory around the primary. This comparison between the halo model A, MOG, and MOND, is shown for a $5~\rm kpc$ luminous sphere with uniform mass distribution in the left panel of Figure \ref{fig:acc_ratio}.
One interesting point is that for the smaller scales, comparable to the non-dark matter scales, the MOND and MOG accelerations are almost equal and behave like the Newtonian case. On larger scales in the range between $10$ and $100~\rm kpc$, the MOG gravitational field is stronger than the others, so that the test particle will be more accelerated.

Here we have a brief description for the asymptotic behavior of the accelerations in MOG and MOND. 
In the gravitational potential of MOG on scales much larger than the length scale of the theory, $\mu^{-1}\approx 24~\rm kpc$, Equation (\ref{eq:potential22}) reduces to the form of a Newtonian gravity enhanced by a factor of $(1+\alpha)$. Thus, in this limit, which we call deep-MOG, the test particle acceleration behaves as $1/r^2$. 
Moreover, according to Equation (\ref{eq:mond}), the acceleration of a test particle in the deep-MOND regime changes as $1/r$.
Defining the characteristic scale of deep-MOND as
\begin{equation}\label{eq:MONDr0} 
r_0 = \sqrt{GM/{a}_0},
\end{equation}
the ratio of acceleration in MOG over MOND is given by $a_{\rm MOG}/a_{\rm MOND} = (1+\alpha){r_0}/{r}$. For $r<(1+\alpha)r_0$, we would therefore expect $a_{\rm MOG}>a_{\rm MOND}$ and for $r>(1+\alpha)r_0$, $a_{\rm MOG}<a_{\rm MOND}$. 
The ratio of the accelerations between MOG and MOND is shown in the right panel of Figure \ref{fig:acc_ratio} for the two uniform spheres with different luminous masses.
We have this $1/r$ property for the ratio of accelerations for $r>30-40~\rm kpc$, depending on the mass of luminous matter. On shorter scales ($r\ll(1+\alpha)r_0$), we have to take into account the full acceleration formula for MOG and MOND.

\subsection{Dwarf Motion}\label{sec:dwarf_motion}

To find the shell radii at any time, one first needs to trace the motion of the incoming galaxy, which is released to move radially toward the host center. This is done by solving the equation of motion of the dwarf and the individual stars under certain conditions. The incoming galaxy moves as a damped oscillator around the center of the primary and releases a part of its mass each time it passes through the center of the host galaxy and forms a new generation of shells.
The oscillations continue until the dwarf loses all its mass or completely merges with the host and the centers of density coincide. 
Multi-generation shell structures, caused by multiple crossings, commonly form in self-consistent simulations~\citep{seguin1996dynamical, bartovskova2011simulations, cooper2011formation}.
It has been shown in test-particle simulations~\citep{dupraz1986shells, ebrova2012quadruple, bilek2013testing} that for a minor merger model, the observed wide range of shell radii requires the shell system to be formed in several generations.
In this study, consistently with the simulations, we choose the dwarf to lose a part of its mass in every passage and all of the remaining mass in the fourth transit, such that after creating four shell generations, dissolution occurs.

\begin{figure}
   \centering
   \includegraphics[width=\columnwidth]{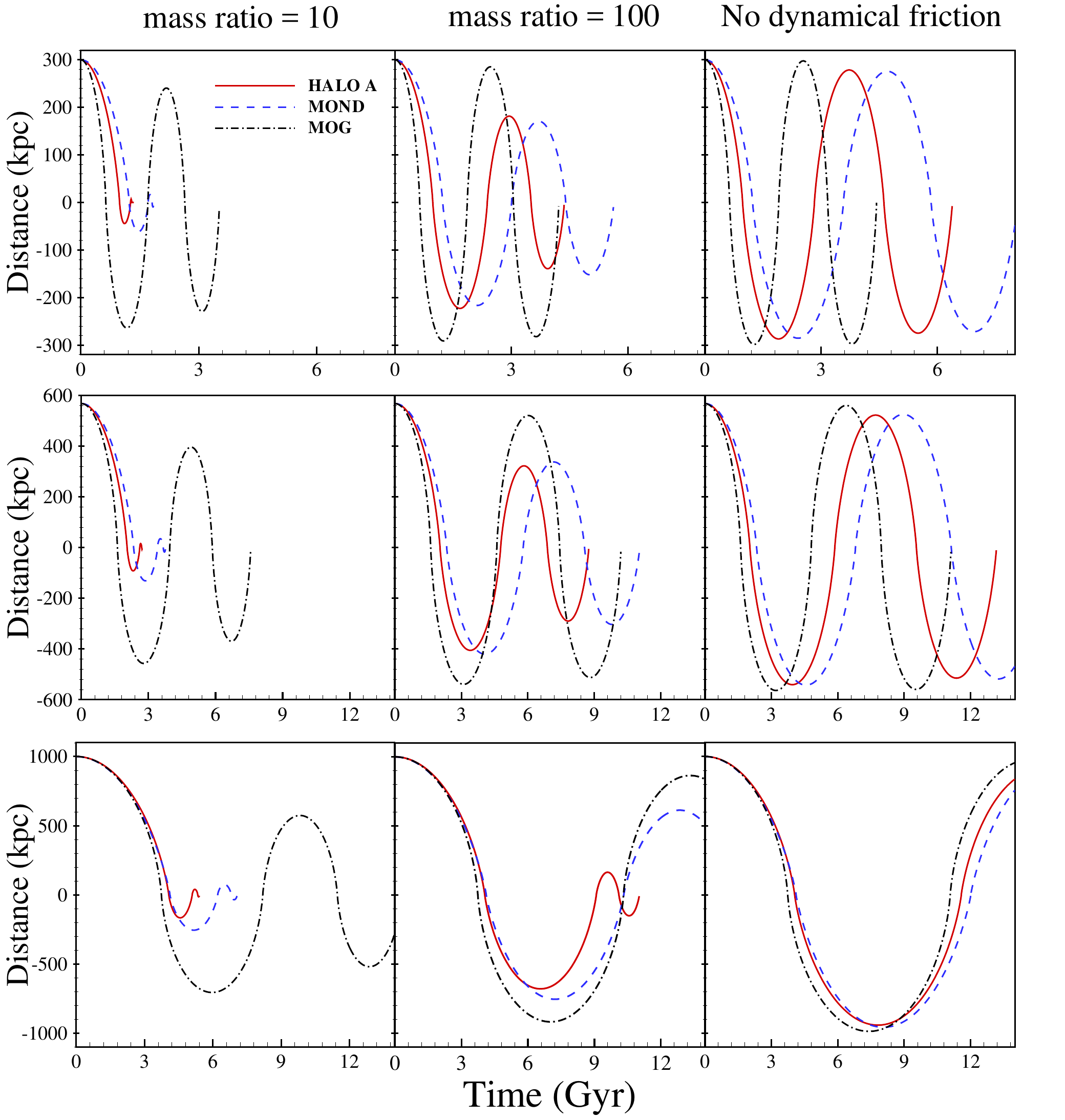}
   \caption{Trajectory of the secondary that completely dissolves into the primary after four passages. Halo model A is shown with the red solid line (see the Table~\ref{tab:models}), MOG with the black dash-dotted line, and MOND with the blue dashed line. The motion starts from rest from inside, the edge, and from outside of the dark matter halo, in the upper, middle, and bottom panels, respectively. $q=10$ refers to the dwarf with dark matter halo and $q=100$ represents the dwarf without dark matter content.
In the last column,  the dwarf mass is completely negligible, such that Chandrasekhar dynamical friction is not operative.}
   \label{fig:core_motion1}
   \end{figure}

\begin{figure}
   \centering 
   \includegraphics[width=1.0\linewidth]{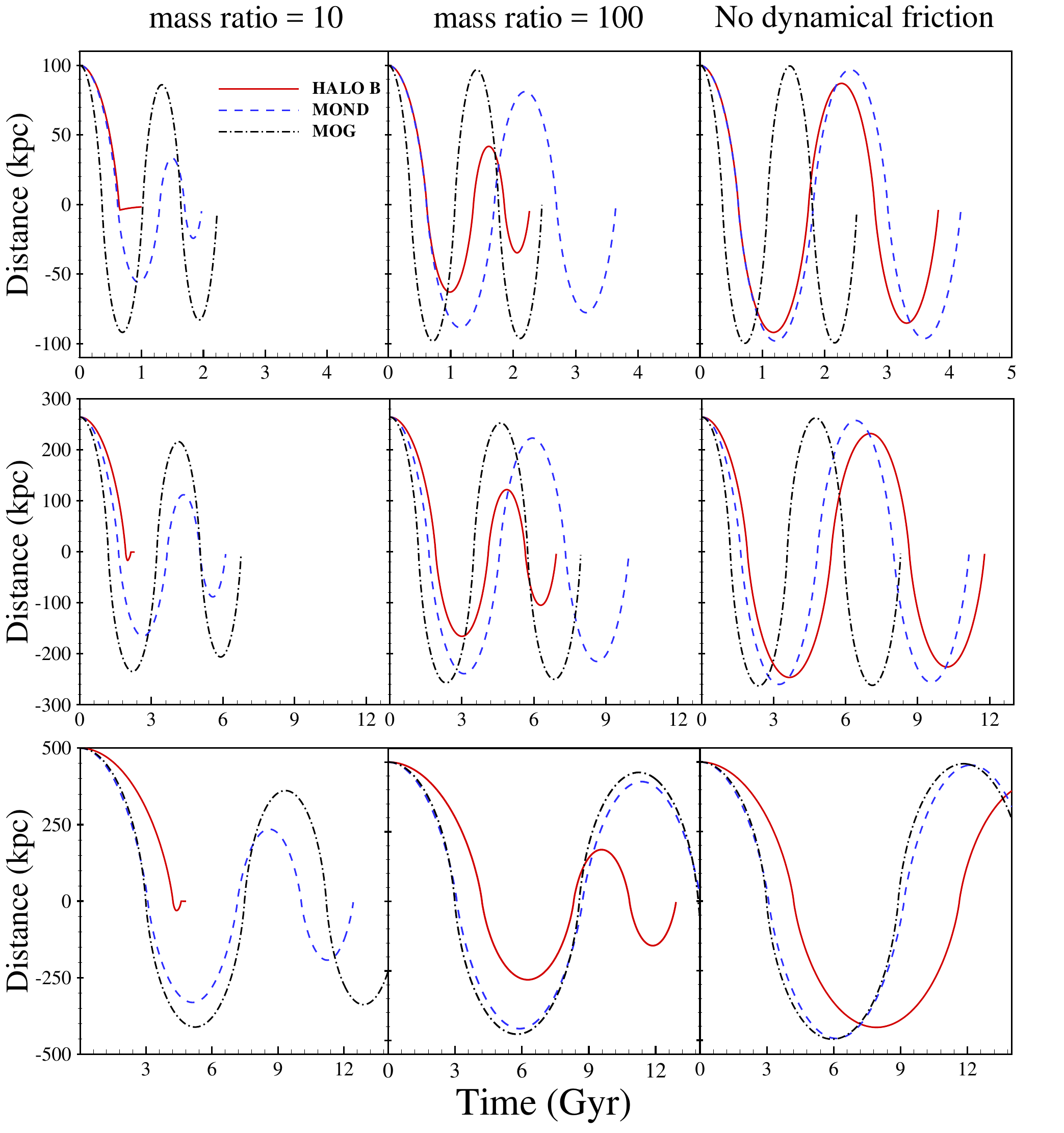}
   \caption{Same as the Figure~\ref{fig:core_motion1}, but with the halo model B instead of the halo model A. We note that the initial distances of the secondary in Figures~\ref{fig:core_motion1} and~\ref{fig:core_motion2} are different.}
           \label{fig:core_motion2}
   \end{figure}

Releasing the secondary at rest from different initial distances, Figures~\ref{fig:core_motion1} and \ref{fig:core_motion2} show the trajectory of the secondary and its core debris after the first passage and compare MOG and MOND with the dark matter halo models A and B, respectively. 
We let the secondary start its motion from three different points: the edge, the inside, and the outside of the dark halo.
Figure~\ref{fig:core_motion1} is related to the halo model A, where the dark mass is 10 times more massive than in the halo model B. In this case, where the virial radius of the halo is equal to $568~\mathrm{kpc}$, the starting points have been chosen to be $300, 568$, and $1000~\mathrm{kpc}$ respectively. For Figure~\ref{fig:core_motion2} these distances take the values of $100, 264$, and $500~\mathrm{kpc}$ and the virial radius of the dark halo is $264~\mathrm{kpc}$.

   
The mass ratio is defined as $q={M^{\ast}_{\rm p}}/{M_{\rm s}}$, where $M^{\ast}_{\rm p}$ is the baryonic mass of the primary and $M_{\rm s}$ is the total initial mass of the secondary.
In the first two columns of the Figures \ref{fig:core_motion1} and \ref{fig:core_motion2} we compared two different cases for all models such that $q$ takes the values of 10 and 100. We also show the results for $q\to \infty$, representing the condition with relatively no dynamical friction in the last columns. 
As a consistency check, it has been shown in the last columns of Figure ~\ref{fig:core_motion1} and \ref{fig:core_motion2} that for a dwarf with sufficiently low mass, the MOG and MOND potentials act the same for large infall distances, as do the halo models. This identical behavior of the models on large scales was expected from comparing the accelerations in different models, which is demonstrated in Figure \ref{fig:acc_ratio} (see section \ref{sec:acc}).
Owing to dynamical friction in different dark matter halo models, having a more massive secondary means greater orbital energy loss within the primary and more slowly released stars, which leads to a brighter system of small shells on a shorter timescale.
 
    \begin{figure*}
   \centering
   \includegraphics[width=1.0\linewidth]{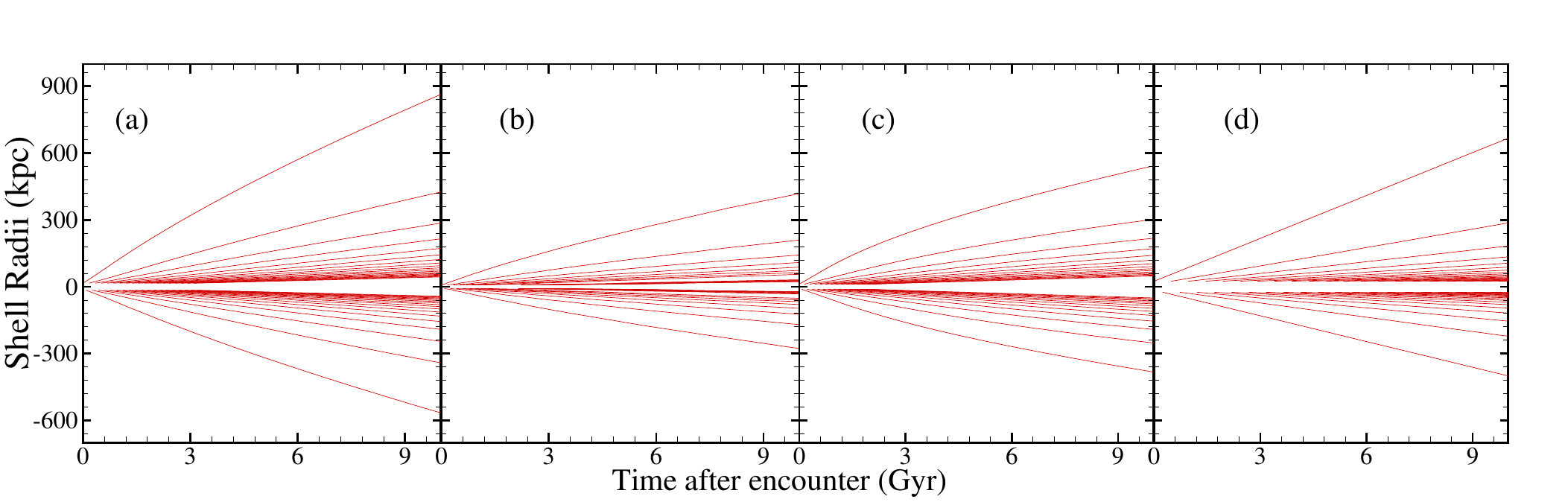}
   \caption{Patterns of shells in the first generation formed in different models of the gravitational potential for the host elliptical. \textbf{(a)} Halo model A, \textbf{(b)} halo model B, \textbf{(c)} MOG, and \textbf{(d)} MOND}     
         \label{fig:shell_pattern} 
   \end{figure*}

When we only consider the red solid curves that represent the dark matter halo models in Figures~\ref{fig:core_motion1} and~\ref{fig:core_motion2}, for different initial conditions, we can investigate the effect of the existence of dark mattar in the dwarf galaxy by comparing the first two columns. 
As the dwarf galaxy is assimilated to a point mass in our study, the only important parameter is the total mass of the dwarf apart from its distribution.
In this case,  we assume the luminous mass of the dwarf galaxy to be the same in both columns, that is, one hundredth of the elliptical galaxy luminous mass. Therefore, we can translate $q=100$ to represent a dwarf without dark matter and $q=10$ to a dwarf with a dark matter halo that its dark mass is 10 times of its luminous mass. 
Comparison between these two cases shows that the presence of dark matter in the dwarf causes faster merging, ending in tighter shell generations. In the case $q=100$ when the second-generation shells appear, the first-generation shells are almost {\it{old}} and {\it wide} and have faded because of the long time between the core passage through the center. 
  
\subsection{Shell Structure}\label{sec:shell_structure}

Shell patterns at any time after an encounter are derived when we follow the oscillation of the released stars and find the positions where they spend more time, as shown in Figure~\ref{fig:shell}. 
The distance between the shell generations is controlled by dynamical friction, which depends on the initial mass ratio between the dwarf and the elliptical and the mass-loss ratio of the secondary in each passage. 
It is clear from Figures~\ref{fig:core_motion1} and~\ref{fig:core_motion2} that in MOND potential (because the dwarf galaxy moves through the stellar body of the host), dynamical friction has a tiny effect on the movement of the secondary core. However it is larger compared to MOG because of smaller relative velocity, and strongly influences the motion in the dark matter halo models. 
It slows down the core so that a massive dwarf sinks into the host quickly and all of the stars generate a bright shell system in only one or two close passages. In the modified gravity models, on the other hand, it takes significantly longer for complete merging to happen, and the shells will be less bright than in the halo models.
Therefore, in the same encounter conditions of the two galaxies, shells appear on a shorter time-scale in the presence of dark matter. 
In this case, the resulting bright shells are mostly composed of low-velocity stars and so appear over a smaller range of distances. Higher velocity stars generate large faint shells. Therefore, the observation of small bright shells should be accompanied by large faint ones because dynamical friction dominates in the halo models.   
In the case of MOG, the velocity of the core remnant after the first passage does not decrease much, and the next shell generations would have a similar velocity distribution as the first. 
The first-generation shells repeat with approximately the same total luminosity and therefore decreasing surface brightness since later shell orders have larger radii. Successive shell generations, however, produce dimmer shells because fewer stars are liberated from the secondary at each passage.
        
Figure \ref{fig:shell_pattern} shows the shell structures in the first-generation shells formed in different models of the gravitational potential. 
We assumed that the stellar velocities take all the possible values from zero to infinity (see Section \ref{sec:shell_br}), therefore the initial conditions of the impact do not play a role in one-generation shell patterns.      
A comparison between the halo models A and B in Figure~\ref{fig:shell_pattern} shows that the size of the shells in the dark matter scenario obviously depends on the amount of dark matter of the primary galaxy, so that this can be used to estimate the total mass of the dark halo (but not the dark matter distribution) in the host galaxy. The appearance times of the the next shell generations in the presence of dark matter depend on the initial mass of the dwarf. From the observational point of view, measuring the integrated surface brightness of the shells would give the approximate initial dwarf mass before the encounter and make the dynamical friction calculation more accurate in case studies.
This figure also shows that there is a degeneracy with time and position in the first shell generation patterns in the halo model B and MOG and MOND.
The exact shell positions therefore cannot be used to verify the models without studying the next shell generations. As we discussed in Section \ref{sec:dwarf_motion}, the distance of the next-generation shells depends on the existence of dark matter. Thus, one has to label a given shell to a specific generation in the observational data. This is not possible unless when the shell ages are defined. Even in this case, without knowledge of the initial distance of the dwarf, MOG and MOND cannot be distinguished from each other because of the negligible dynamical friction.

\begin{figure}
   \centering
   \includegraphics[width=1.06\linewidth]{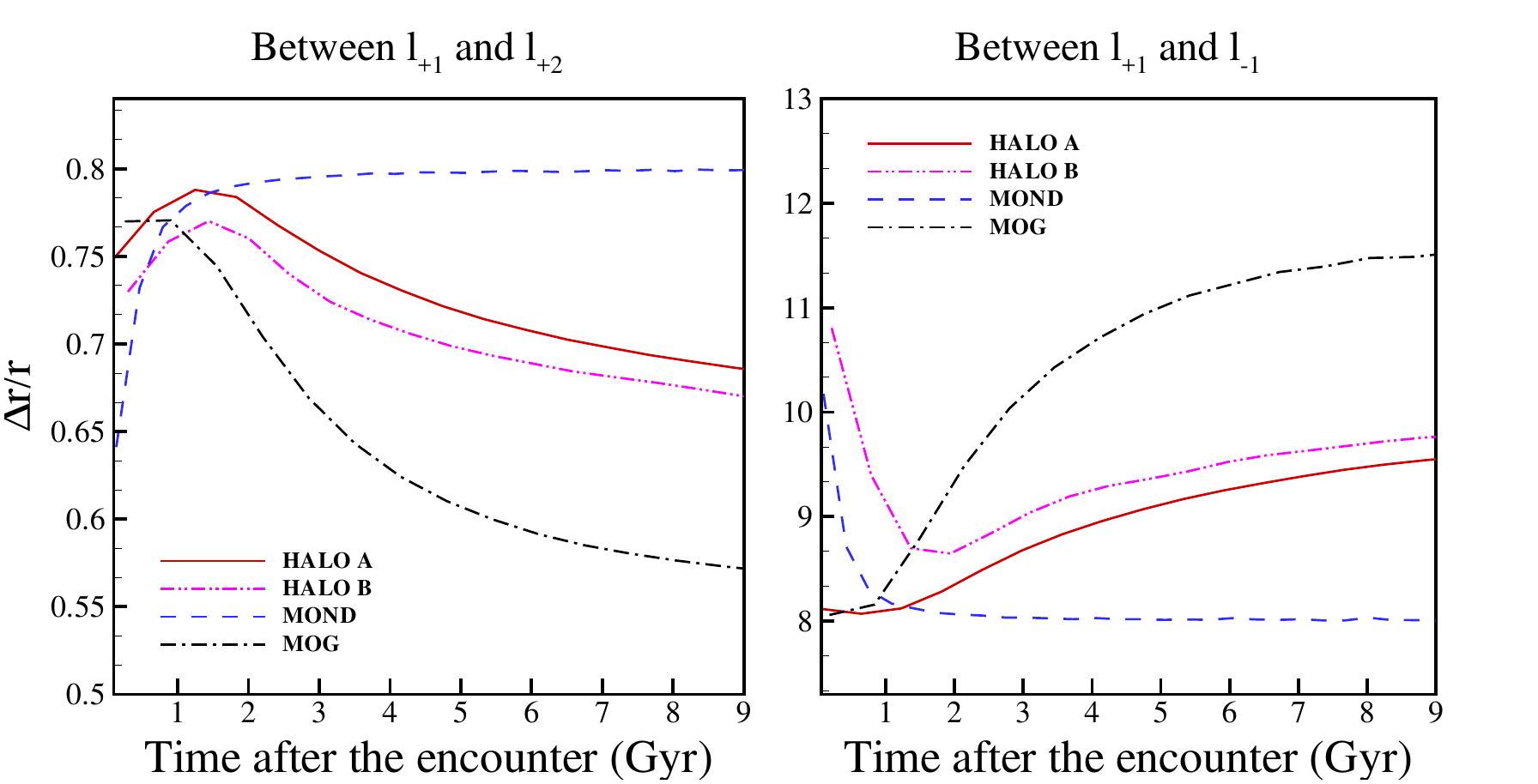}
   \caption{Time evolution of the relative distance of shells (Equation~\ref{eq:delta_r}) in different models. Halo model A is shown with the red solid line, halo model B with the purple dash-double-dotted line, MOG with the black dash-dotted line, and MOND with the blue dashed line. Left: the first outer shells in the same direction ($l_{+1}$ and $l_{+2}$). Right: the outer shells in the opposite direction ($l_{+1}$ and $l_{-1}$).}
   \label{fig:deltar_r} 
    \end{figure}

To find a criterion to more easily distinguish between the models, one needs to employ the dynamics of the shells as derived from the equation of motion of the stars in different potentials.
The scale-dependent variation of the accelerations in the different models leads to different dynamics of the shells (see Section \ref{sec:acc}).
We compare the evolution of the relative distances of the shells in our models defined as
\begin{equation}\label{eq:delta_r}
\frac{\Delta r}{\bar{r}}=2 \left( \frac{r-r^{\prime} }{r+r^{\prime}} \right),
\end{equation}
where $r$ and $r^{\prime}$ are the shell radii from the center of the primary galaxy.
According to Figure \ref{fig:shell} the outermost shell ($l_{+1}$) always contains the stars that have completed three-fourths of their period. The stars in the second shell in the same direction ($l_{+2}$) have passed seven-fourths of their oscillation period. The shell $l_{-1}$ is the first on the opposite side from the originally incoming dwarf.
We show the time evolution of ${\Delta r}/{\bar{r}}$ for the shells in different models in Figure~\ref{fig:deltar_r} between $l_{+1}$ and $l_{+2}$ in the left panel and between $l_{+1}$ and $l_{-1}$ in the right panel. 
As we can see in Figure~\ref{fig:deltar_r}, the relative distances are significantly different between the different gravitational theories for times longer than $2~\mathrm{Gyr}$ after the first encounter.
Thus, measuring the relative distances of the outer shells on the two sides of the galaxy long enough after the encounter is a potentially powerful tool to decide which gravity theory has governed the merger process. In addition, among the modified gravity theories, MOG and MOND can be distinguished from each other via this probe.
Being dimensionless and independent of the distance of the galaxy is the advantage of this probe, in comparison to the position of the shells.

\begin{figure}
   \centering
   \includegraphics[width=1.07\linewidth]{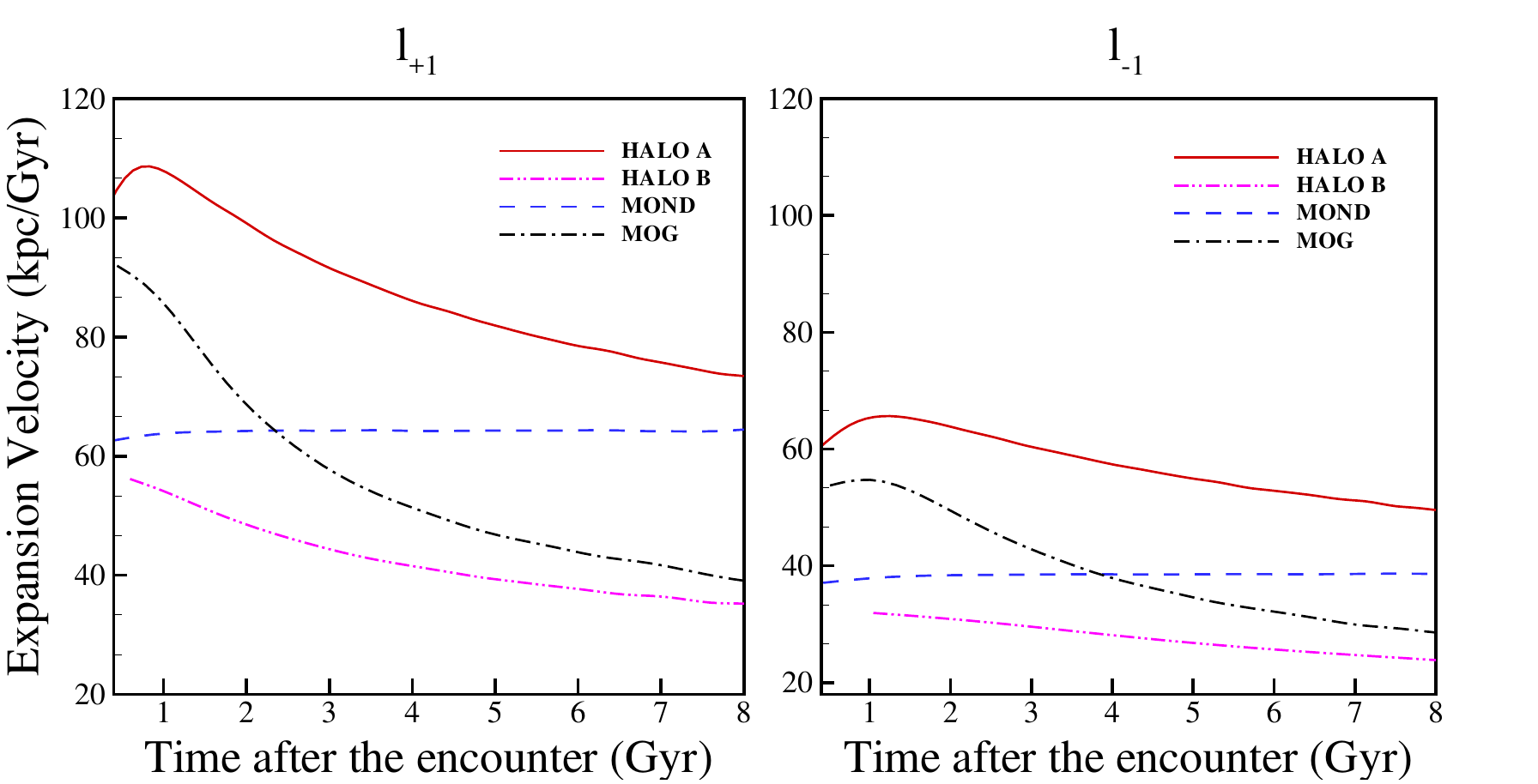}
   \caption{Model-dependent behavior of the shell expansion velocity. Lines are the same as Figure \ref{fig:deltar_r}. Left: The outermost shell ($l_{+1}$). Right: The first shell in the opposite direction ($l_{-1}$).}
   \label{fig:exp_vel} 
       \end{figure}

According to Equation (\ref{eq:exp_vel}), the expansion (phase) velocity of the shells can also be used as a probe to verify the existence of dark matter or to decide whether it is necessary to modify the theory of gravity. Figure~\ref{fig:exp_vel} depicts this quantity for $l_{+1}$ and $l_{-1}$ for the first-generation shells. It shows that an almost constant shell expansion velocity is the {\it{property}} of MOND, while in MOG and the dark matter halo models, the expansion velocity decreases with time.
Moreover, the more massive the dark matter halo surrounding the galaxy, the faster the expansion of the shells. 
 
As a practical example, we take the case of NGC 3923, a unique Type I shell galaxy, which is surrounded by many stellar shells~\citep{bilek2016deep}.
There are 18 redshift-independent distances in the NASA/IPAC extragalactic database for this galaxy, with a median of $20.8~\mathrm{Mpc}$.\footnote{\url{http://ned.ipac.caltech.edu/cgi-bin/nDistance?name=NGC+3923}} If we take the typical expansion velocity of the outermost shell for this galaxy to be $\approx100~\mathrm{kpc~Gyr^{-1}}$, the metrical displacement rate of the shell would be $\Delta \Theta / \Delta t \approx1~\mathrm{mas~yr^{-1}} $. 
The typical evolution of velocity for this shell would be $\Delta^2 \Theta / \Delta \mathrm{t}^{2} \approx 10^{-11}\mathrm{\mu as~yr^{-2}} $.
~\cite{ebrova2012quadruple} developed the usage of the shell spectral line profile to measure the expansion velocity of the shell. 
They used an analytical approach and test-particle simulations to predict the line-of-sight velocity profile across the shell structure and showed that
spectral peaks are split into two, giving a quadruple-peaked line profile that enables us to directly measure the shell expansion velocity.
           
\begin{figure}
   \centering
   \includegraphics[width=1.05\linewidth]{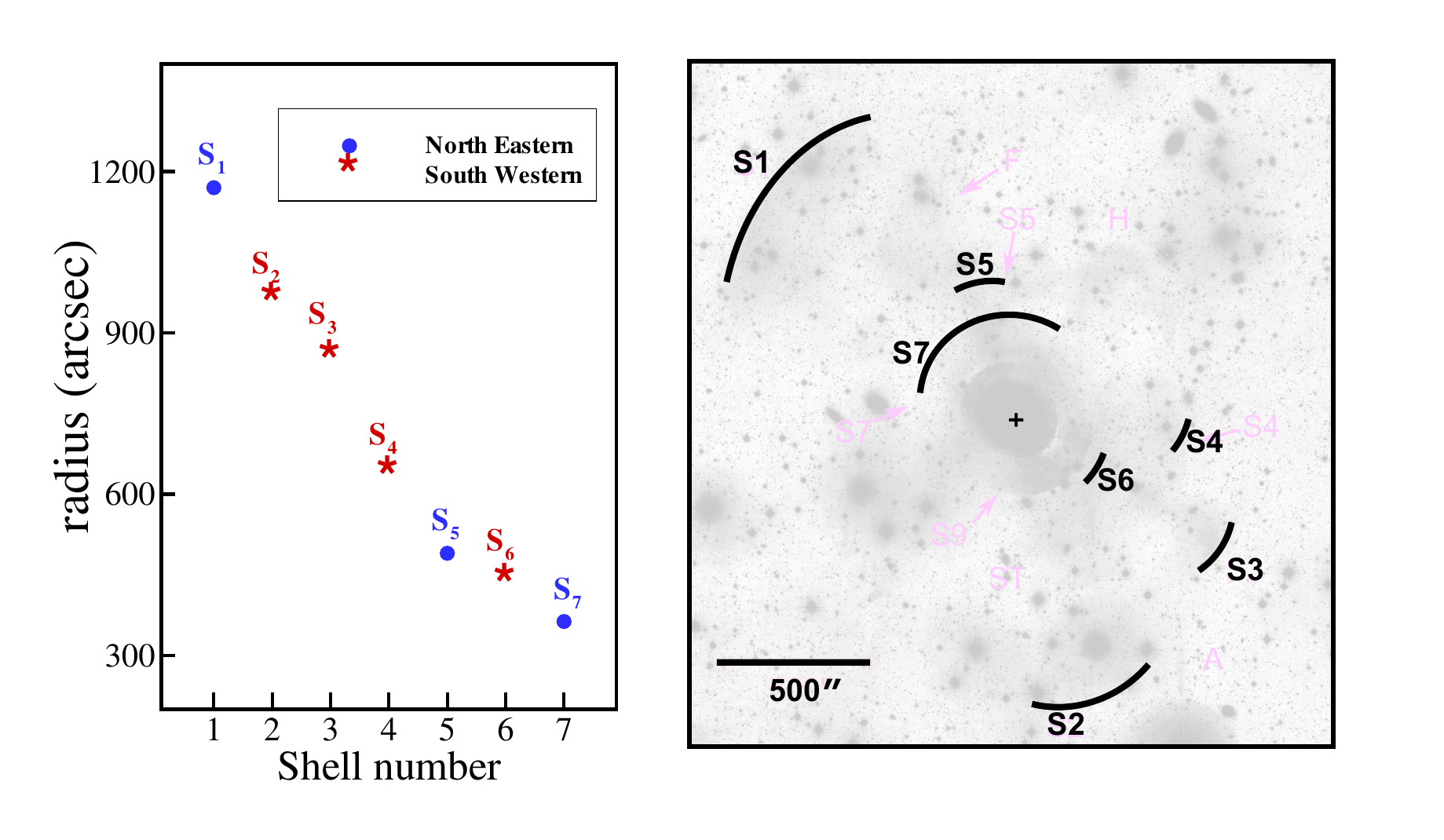}
   \caption{First seven observed shells of NGC 3923. Data and the underlying photo of the galaxy is adopted from~\citet{bilek2016deep}. Left: distances from the center of the galaxy. Blue dots represent the northeastern shells, and red star points show the southwestern shells. Right: spatial positioning of the shells around the galaxy. 
   The labels $\rm S_i$ are explained in Table \ref{tab:data}. For a distance of $20.8~\rm kpc$, $500^{\prime\prime}$ corresponds to $49.92~\rm kpc$.}
   \label{fig:data} 
       \end{figure}

To compare the model predictions relative distances of the shells (Equation \ref{eq:delta_r}) with observational data, we use the data of \cite{bilek2016deep}. They observed 42 shells around NGC 3923 with Megacam, which is the highest number of shells amongst shell galaxies. The summary of their data for the spatially largest shells is given in Table \ref{tab:data} and Figure \ref{fig:data}.
The luminous mass of the halo model A is $5 \times 10^{11}{ M_{\odot}}$ (Table \ref{tab:models}), which is consistent with the value reported for NGC 3923 by~\cite{bilek2013testing}. Halo model A should therefore be valid for the mass profile when we have dark matter.

According to the merger theory for shell formation, the shells should be interleaved in radius. Thus, a quick look at the data reveals that there must be some missing shells between $\rm S_2$ and $\rm S_3$ and also $\rm S_3$ and $\rm S_4$ in the northeastern side of the galaxy. Any further analysis for the largest shells would be possible after observing the missing shells. However, $\rm S_2$ and $\rm S_4$ are possibly uncertain shells according to their observational prominence, as are shells $\rm S_5$ and $\rm S_6$.
Therefore we consider only the most prominent data, \textit{i.e.} $\rm S_1$, $\rm S_3$, and $\rm S_7$, as the certain shells, as if there were no shells in the other reported positions. That is we discard shells with prominence degrees 3 and 4.

\begin{figure}
   \centering
   \includegraphics[width=1.06\linewidth]{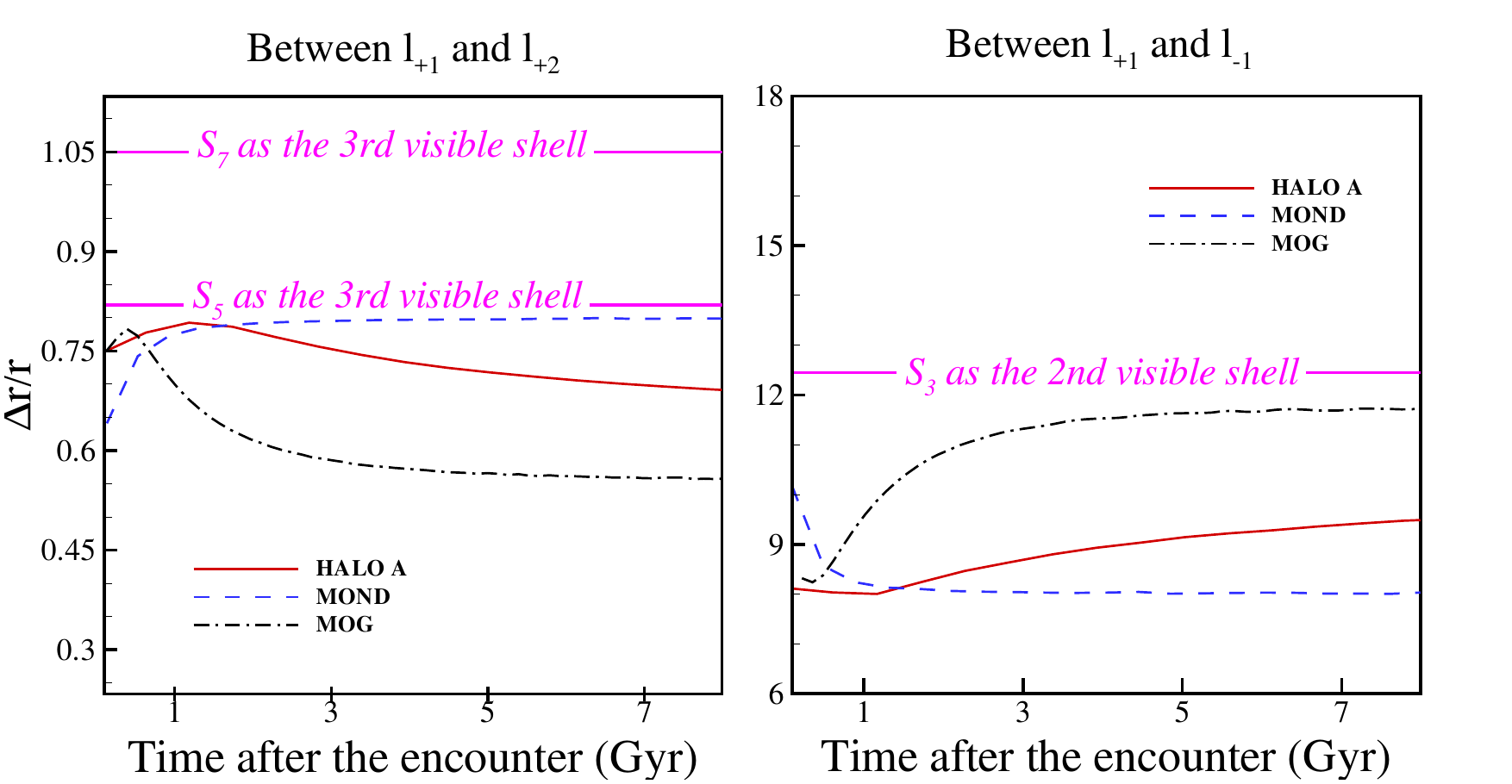}

   \caption{Comparing the data of shells of NGC 3923 reported by~\citet{bilek2016deep} with the predicted curves of the relative distances of the same-generation shells in different gravity theories. The panel ordering and the model lines are the same as in Figure \ref{fig:deltar_r}. The purple lines show the observational values assuming $\rm S_1$ is the largest visible shell. 
      }
   \label{fig:dr_r_obs} 
       \end{figure}

Figure \ref{fig:dr_r_obs} shows the model comparison with the data. The model relative distances are calculated for the first-generation shells.
In the right panel the relative distance of the largest opposite shells ($\rm S_1$ and $\rm S_3$) are plotted. In this case,  MOG is more consistent with the data. The left panel shows the relative distance of the two outermost successive shells. The radial distance of $\rm S_1$ and $\rm S_7$ is very large such that it is possible for $\rm S_7$ to belong to another shell generation. Thus, inconsistency with the models is possible. 
A possibility is that there should be an unobserved shell at a distance larger than $\rm S_7$ in the northeastern side of the galaxy.
As a possible candidate for this shell we considered $\rm S_5$ and checked the consistency in the models and the data. MOND is more consistent for this case.
However, for a precise comparison, one should have detailed \textit{N}-body simulations for each model, deliberating the characteristics of this special galaxy.
 
  \begin{deluxetable}{cccc} 
\tablecaption{\label{tab:data}Summary of the Data of the First Seven Shells observed by \citet{bilek2016deep}, Taken from Their Table 2.}	

\tablehead{
\colhead{Label} & \colhead{Radius} & \colhead{Prominence} &  \colhead{ Note} \\
\colhead{} & \colhead{(arcsec)} & \colhead{} & \colhead{} 
}
\startdata
	$\rm S_1$ & +1170 & 2 & largest \\
	$\rm S_2$ & -952 & 4 & highly uncertain \\
	$\rm S_3$ & -846 & 1 & narrow \\
	$\rm S_4$ & -630 & 3 & irregular, diffuse \\
	$\rm S_5$ & +490 & 3 & possibly not a shell \\
	$\rm S_6$ & -430 & 4 &  highly uncertain \\
	$\rm S_7$ & +363 & 1&  ...\\
\enddata
	
\tablecomments{Shells lying on the northeastern side of the galaxy have a positive sign; the shells on the southwestern side of the galaxy have a negative sign. The prominence degree describes the shell detection certainty as 1 = prominent and sharp edged, 2 = prominent and diffuse edged, 3 = faint but probably existent, and 4 = questionable.}
\end{deluxetable}

\section{Conclusions} \label{sec:conclusion}

We have solved the two-body problem of the radial collision of a dwarf galaxy with an elliptical galaxy until complete cannibalization with the elliptical in different gravity theories for identical initial systems. During this collision, we have studied the shell formation process in order to emphasize shell galaxies as an observational tool for testing gravity models.
We have compared two dark matter halo models versus MOG and MOND as modified gravity models without dark matter (see Table~\ref{tab:models}).
We showed that theoretically, there exist parameters based on shell observations that can be used to distinguish between different gravity theories. We presented a schematic pattern of the shell structure, including the shell positions and apparent shell expansion, in Figure~\ref{fig:shell}.

The integrated surface brightness of the resulting shell pattern and the distance between the shell generations are observables that place limits on the initial conditions of the merger. 
Given constraints on the initial condition, the gravitational potential of the host galaxy can be uniquely defined by the shell distribution and their brightness.

Our study of the motion of the dwarf inside the elliptical has shown that Chandrasekhar dynamical friction strongly affects the merger process in the dark matter halo models, resulting in compact bright shell generations soon after the encounter.
On the other hand, in MOG and MOND potentials, in which dynamical friction is smaller, the shells are dimmer and the time intervals between generations are longer.
When the dwarf mass is negligible such that dynamical friction is not operative, the motion in MOG and MOND and in halo model A are almost the same on large scales, while the halo model B behaves significantly differently (Figures~\ref{fig:core_motion1} and \ref{fig:core_motion2}).

In the dark matter halo models, the presence of dark matter in the dwarf makes tighter shell generations, while in the absence of dark matter in the dwarf, the first-generation shells are {\it{old}} and {\it wide} when the second-generation shells appear.
In general, we conclude that the observation of small bright shells accompanied by large faint shells could be regarded as a sign of the existence of dark matter in and around the elliptical.
In the modified gravity theories, we expect the corresponding shells in terms of patterns to have a self-similarity because the encounter velocity of dwarf-elliptical does not decrease much since dynamical friction is smaller.
It should be noted that in all cases, next-generation shells are commonly fainter because fewer stars are releases at each passage.   
We also emphasize that the merging times may be too rapid if galaxies have dark matter haloes (Figures~\ref{fig:core_motion1} and \ref{fig:core_motion2}) in comparison to the need for the dwarf to oscillate more than one time to form multi-generation shells, as suggested by the wide range of shell radii in observations~(see Section \ref{sec:dwarf_motion}).

Since there is a degeneracy between the shell pattern with the age of the shell system in one generation, the initial conditions, and the gravity models, we have introduced the relative distance of the first-generation shells (Equation \ref{eq:delta_r}) to represent differences between different models. 
We conclude from Figure~\ref{fig:deltar_r} that the relative distance of the first-generation shells on both sides of the galaxy is the observable parameter to break the degeneracy and to reveal the underlying gravity theory sufficiently long after the encounter. The shell expansion velocity (Figure \ref{fig:exp_vel}) is another observable to clarify the model differences. 
We note that the shell expansion velocity decreases in time in all of the models except for MOND.
While there is dynamical degeneracy between the dark matter halo models and MOG for spiral and elliptical galaxies and clusters of galaxies~\citep{haghighi2016testing}, the phase velocity of the shells can break this degeneracy in shell galaxies.
Consistency of the model predictions for the relative distances of the shells with the data for NGC 3923, considering the uncertainties of our models, has been investigated in Figure \ref{fig:dr_r_obs}.

Theoretical assumptions that we used for the encounter include a purely radial collision and the procedure for liberating stars from the dwarf at the passage through the center of the elliptical. Although these assumptions have already been made in previous theoretical studies (e.g. \citealt{hernquist1987shell}), test particle simulations showed deviations from the results assuming analytical radial motions of the stars and releasing the particles by switching off the dwarf potential exactly at the time of passage. This causes a smearing in the final shell pattern~\citep{ebrova2012quadruple}. 
However, simulations with different spherical potentials by \cite{bilek2016galaxy} showed that the shell radii depend only on the potential of the primary and probably on where the secondary disrupts.
In addition, departures from a radial collision can change the dynamical friction effect and modify our analysis~\citep{seguin1996dynamical}.
Additionally, in modeling the gravity of the host galaxy, we used simplified assumptions. 
We assumed a spherically symmetric host galaxy and so used the full-MONDian acceleration field and ignored the external field effect in MOND.
Moreover, computations of the acceleration in MOG have been accomplished in the weak-field limit of the theory.
To advance the analysis of comparing different gravitational theories with observation and to quantify detailed differences, \textit{N}-body simulations within all gravitational models with identical conditions are required.

\begin{acknowledgements}
 H.V. thanks the Argelander Institute for Astronomy at the University of Bonn, especially P. Kroupa, for hospitality during the preparation of this work.
\end{acknowledgements}

\appendix

\section{Dynamical friction}\label{App:dynamical_friction}

When a massive particle moves through a large system of much lighter particles, it gradually slows down as the exchange of energy-momentum that is due to the many gravitational encounters acts like a friction force on it.
The long-range stronger gravity in modified gravity theories makes the same motion at a certain distance from the baryonic mass of the host, and to a certain degree, mimicks the presence of a dark matter halo.
Meanwhile, the dynamical friction effect efficiently slows down the moving particle as a local force. At long distances from the luminous part of the host, it is therefore reasonable to take dynamical friction as being non-operative in MOG and MOND.
This work deals with dynamical friction of the satellite as it moves through the dark matter halo of the elliptical in halo models A and B, and also as it moves within the stellar body of the primary in all galaxy models. 

The nonlinear nature of MOND's field equations makes dynamical friction calculations in this model non-trivial. \citet{ciotti2004two} calculated the two-body relaxation time and showed that it is shorter in MOND than in an equivalent Newtonian system, by the square of the factor by which the gravity is enhanced. Their conclusion was confirmed by \citet{sanchez2006extensive}. Simulations by \citet{nipoti2008dynamical} also showed that for an \textit{N}-body system dynamical friction within the stellar body of a galaxy is significantly stronger in MOND.
Furthermore, the dynamical friction in MOG acts more intensely as the gravitational force is stronger than in an equal Newtonian system with the same mass.

However, in our calculations in the non-Newtonian models, the incoming galaxy is in the Newtonian regime when it moves within the stellar body of the host, therefore one can approximately use the Newtonian Chandrasekhar formula for the dynamical friction \citep{2008gady.book.....B}. The validity of the Chandrasekhar formula in galactic dynamics is investigated e.g. by~\cite{cora1997orbital},~\cite{verdes2010galaxies}, and~\cite{fellhauer2000superbox}. This is also relevant for the case of MOG, in which the scale of the theory, $\mu^{-1} \approx 24~\mathrm{kpc}$, is much larger than the size of the model galaxy.

If the velocities of the stars in the host have a Maxwellian distribution, the Chandrasekhar formula for the deceleration of the primary, due to dynamical friction from the primary's stars, with mass $m$ and velocity $\textbf{\textit{v}}$ is 
\begin{eqnarray}\label{eq:df-iso}
\frac{{\rm d} {\textbf{\textit{v}}} }{ {\rm d}t} &=&  \Gamma_{\rm df} \, \textbf{\textit{v}},
\\
\Gamma_{\rm df} &\equiv &
- \frac{4 \pi G^2  \rho m}{v^3}\ln \Lambda \left[ {\rm Erf}(X) - \frac{2 X}{\sqrt{\pi}}\, {\rm e}^{-X^2} \right], \nonumber
\end{eqnarray}
where $\rho$ is the density of the host, $X\equiv v/(\sqrt{2}\sigma)$ and ${\rm Erf}(x)=(2/{\sqrt{\pi}})\int_0^x {\rm e}^{-y^2} {\rm d} y$ is the error function. $\sigma$ is the velocity dispersion, which for a virialized system of particles with isotropic velocity distribution can be estimated as $ \sigma^2   \approx  {(G M)}/{(5R)}$, where $M$ is the mass and $R$ is the characteristic radius of the host.
The quantity $\ln \Lambda$ is the {\sl Coulomb logarithm}, and its value depends on the problem at hand. 
For radial motion in Newtonian gravity, we use $\ln \Lambda = \ln [(3 M)/(5m)]$~\citep{aceves2007introduction}.




\bibliographystyle{aasjournal} 
\bibliography{1.bib} 




\end{document}